\documentclass[
    ,final            
 ,numberedheadings 
  ]
  {aipproc}

\layoutstyle{6x9}

\begin{document}

\title{Constraining Binary Evolution with Gravitational Wave Measurements of
Chirp Masses}

\author{Tomasz Bulik}{
  address={Nicolaus Copernicus Astronomical Center, Bartycka 18, 00716 Warsaw,
  Poland}
}

\author{Krzysztof Belczy\'nski}{
  address={Northwestern University, Dept. of Physics \& Astronomy,
       2145 Sheridan Rd., Evanston, IL 60208, USA}
}
 
\begin{abstract}
 Using the StarTrack binary population synthesis code we
investigate the properties of population 
of   compact object binaries. Taking into account the  
selection effects we calculate the expected properties
of the observed binaries.We analyze possible constraints 
on the stellar evolution models and find that an observed sample
of about one hundred mergers will yield strong constraints on the  
binary evolution scenarios.
\end{abstract}

\maketitle

\section{Introduction}
 
During this meeting we have learned about the great  progress in gravitational
wave  astronomy. LIGO has  finished its first two data runs and the data
analysis  is being done. The sensitivity of this instrument is improving
steadily. The volume of space in which  LIGO is sensitive to sources of
gravitational wave radiation is  increasing.   While the main challenge is
still  detection of gravitational waves to prove directly their existence,
another set of question arises: once we see sources of gravitational waves 
what astrophysical significance shall they have? Can any   astrophysical
problems be solved  with gravitational wave astronomy? In this  paper we
concentrate on the most promising sources  of high frequency gravitational
waves (at least in the opinion of the authors), i.e. on coalescences of 
compact object binaries. We know that such binaries exist and that all the
observations in the electromagnetic domain are consistent with emission of
gravitational  waves by them. We also know that these objects will
coalesce.

So far most of the work on such binaries in the context of gravitational wave
observations has concentrated on calculating the expected rates  for the
interferometric detectors. This  problem has been  approached in two ways. The
first approach was based upon studying and analyzing the known compact object
binaries, and then considering selection effects to estimate  the properties of
the entire population of such sources to finally obtain the coalescence rate.
The drawback of this approach is the small number statistics, or even  zero
object statistics, in the case of black hole neutron star or double black hole
binaries. Moreover, the results have also suffered from the  uncertainty in
estimates of the selection effects 
\citep{1991ApJ...379L..17N,2001ApJ...556..340K}.  A second approach is based on
studying the stellar evolution processes and detailed analysis of the formation
paths of double compact object binaries
\citep{1997MNRAS.288..245L,1998ApJ...496..333F,1998A&A...332..173P,1998ApJ...506..780B,1999ApJ...526..152F,1999MNRAS.309..629B}.
The main problem associated with these calculations is lack of detailed
knowledge of the physics of several important  stages in the stellar
evolution. The results depend strongly on a particular parametrization   of
such stages. For example the formation rates of double neutron star
binaries strongly depend on the distribution of kick velocities  newly nascent
neutron stars receive  in   supernova explosions.

We have studied a different aspect of   observations of gravitational waves
from coalescing binaries \citep{2003ApJ...589L..37B} - a measurement of the
chirp mass. This paper is an expansion of the already published results.
Previously  \citep{2003ApJ...589L..37B} we only considered the  case of
Euclidean space with  constant star formation rate and here we present  a
consideration of   cosmological effects. In section 2 we present the stellar
population model,  section 3 contains  estimates of the constraints from chirp
mass measurements, and we finish with conclusions in section 4.

\section{Calculations}

\begin{table}
\caption{Description of different population synthesis models
used here.}

\begin{tabular}{lp{10.7cm}}
\hline \hline
Model & Description \\
\hline
A      & standard model described in  Belczynski, Kalogera, Bulik (2002),
        but with $T_{Hubble}=15$Gyrs \\
B1,7,11 & zero kicks, single Maxwellian with
$\sigma=50,500,$\,km\,s$^{-1}$, \\
B13 &   Paczynski (1990) kick with $V_k=600$km\,s$^{-1}$\\
C      & no hyper--critical accretion onto NS/BH in CEs \\
E1--3  & $\alpha_{\rm CE}\times\lambda = 0.1, 0.5, 2$ \\
F1--2  & mass fraction accreted: f$_{\rm a}=0.1, 1$ \\
G1--2  & wind changed by\ $f_{\rm wind}=0.5, 2$ \\
J      & primary mass: $\propto M_1^{-2.35}$ \\
L1--2  & angular momentum of material lost in MT: $j=0.5, 2.0$\\
M1--2  & initial mass ratio distribution: $\Phi(q) \propto q^{-2.7}, q^{3}$\\
O      & partial fall back for $5.0 < M_{\rm CO} < 14.0 \,M_\odot$\\
S      & all systems formed in circular orbits\\
Z1--2  & metallicity: $Z=0.01$, and $Z=0.0001$\\

\end{tabular}

\end{table}

We use the StarTrack binary population synthesis code 
\citep{2002ApJ...572..407B}, to analyze formation channels and properties of
  compact object binaries. Within the code
the single stellar evolution is described using a set 
of analytical   formulae \citep{2000MNRAS.315..543H}.
We include the description of the main sequence stars, evolution on the
Hertzsprung gap, the red giant branch, core helium burning phase,
and the asymptotic giant branch. Moreover, we include a detailed 
description of helium stars on their main sequence and giant branch.
The binary evolution description includes such processes as 
orbital changes due to wind mass loss, tidal interaction, and magnetic breaking.
We allow for different models of mass transfers: conservative, 
quasi dynamic, common envelope evolution, and we investigate possibility of
hyper accretion onto compact objects. Description of the supernovae
explosions include the consideration of various kick velocity distributions
and results of numerical modeling of the formation of compact objects 
\citep{1999ApJ...522..413F}. We assume 
that the distributions of the initial parameters are independent: 
the mass of the primary is drawn from a $\propto M^{-2.7}$ distribution 
\citep{1986FCPh...11....1S}, the mass ratio comes from  a flat distribution,
 the distribution of eccentricity is $\propto e$, and the orbital separation 
 distribution is flat in $\log a$.
 In order to test the sensitivity of the results we run the code
varying several parameters. We denote  our standard set of parameters
as  model  A. Descriptions of other models used in this paper
is listed in Table~1. \nocite{1990ApJ...348..485P}

\begin{figure}
\includegraphics[width=0.8\textwidth]{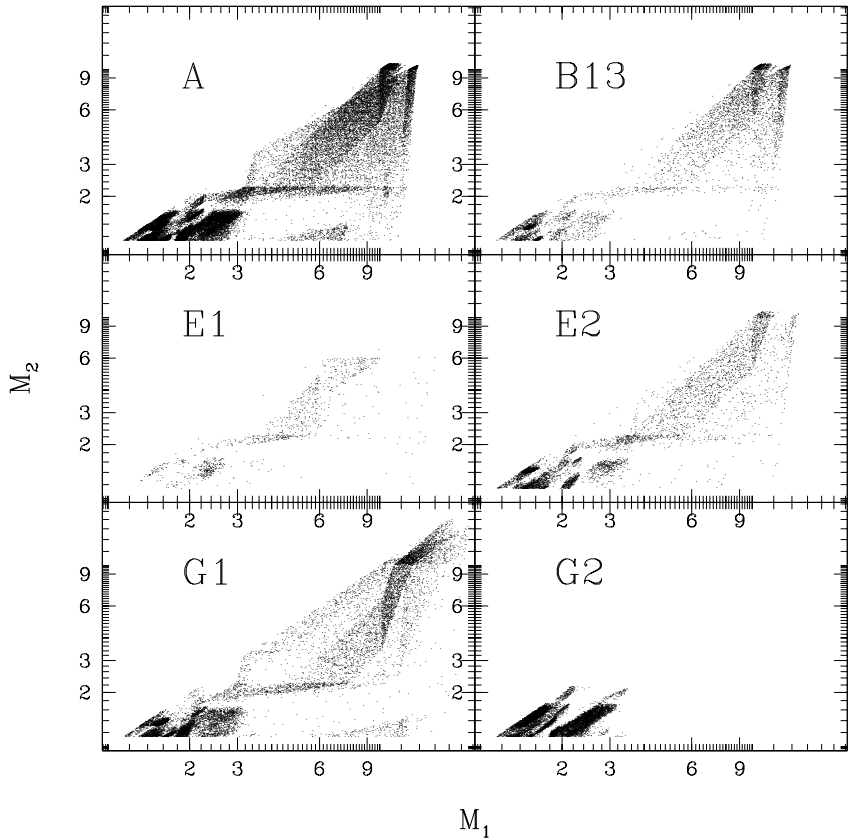}
\caption{Tye distributions of masses of the population of 
compact object binaries in different models of stellar evolution,
for description  see Table~1.}
\label{mmdot}
\end{figure}

A typical calculation involves a few million initial binaries
and results in several thousand compact object binaries.
For each binary we note the masses of the individual   objects 
and the orbital parameters.  The distribution of the masses
of the compact objects appears to be a very sensitive function of the model
of stellar evolution.
We present a few representative examples of such distributions
in Figure~\ref{mmdot}. The models presented are model A,
model B13 with a different kick velocity distribution,
models E1 and E3 where we vary the efficiency of common envelope 
evolution and models G1 and G2 with decreased and increased stellar winds,
respectively. The maximum masses
of the compact object binaries vary from one model to another.
In model G1 with decreased stellar winds there is a large number of 
double black hole binaries with masses up to $20\,M_\odot$ which
are absent in other models. In model E1 there are very little high mass ratio
binaries, and the number of light (double neutron star)
binaries is small compared to double black hole ones. Each model
leads to a different pattern in the mass distribution which will affect
predictions for the gravitational wave observations.

\begin{figure}
\includegraphics[width=0.9\textwidth]{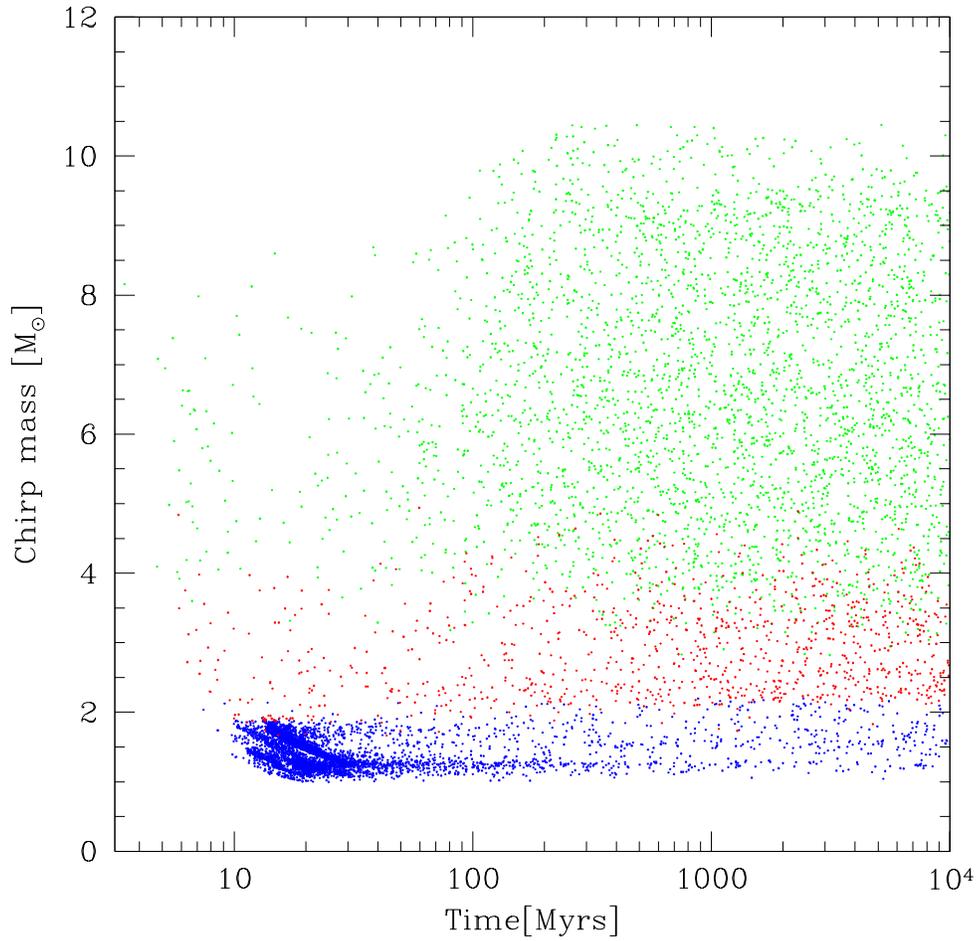}
\caption{Population of the double compact objects within the framework
of model A plotted in the space spanned by the chirp mass, and 
lifetimes counted from the formation of the system on zero age main
sequence to the merger due to gravitational wave emission. The cutoff 
on the left side is due to nuclear evolution time.}
\label{chte}
\end{figure}

A fundamental quantity considered in calculating gravitational waveforms
for coalescing binaries is the chirp mass ${\cal M}= (m_1 m_2) ^{0.6}
(m_1+m_2)^{-0.2}$, where $m_1$
and $m_2$ are the individual masses of the binary components.
In Figure~\ref{chte} we plot the population of binaries from model A
in the space spanned by the chirp mass and the binary lifetime $t_{life}$,
defined as the duration from the formation of a system on zero age main sequence
until its merger due to gravitational wave emission.
A clear tendency is visible in the graph. The low chirp mass
binaries, tend to merge predominantly on a short timescale, on the order of
a few tens of million years. The high chirp mass binaries
have longer lifetimes: a majority of them lives for a few Gyrs.
This leads to an interesting conclusion: the currently merging 
double neutron stars originate in recent star formation episodes, while
the the progenitors of  currently merging  double black holes 
date to a few billion years ago. Black hole neutron star binaries
have  chirp masses between $2$ and $4\,M_\odot$, and  
the distributions of their lifetimes 
is practically flat between a few tens of million years and the Hubble time.

\section{Gravitational wave observations}

We consider here detection of compact object coalescence 
in the inspiral  phase and neglect the merger and ringdown signal.
The signal to noise ratio from an inspiraling binary 
is proportional to $\propto [(1+z){\cal M}]^{5/6}/d$ 
\citep{1993ApJ...411L...5C,1994Bon,1998PhRvD..57.4535F},
where $d$ is the luminosity distance to the source, and
$z$ is the redshift. 
For the stellar mass sources with chirp masses 
below $20\,M_\odot$ the signal to noise from the merger
and ringdown phases is weaker than the inspiral signal
\citep{1998PhRvD..57.4535F}. For a given, assumed signal to noise
value required for a detection of a gravitational wave sources
one can calculate the sampling distance (or redshift), 
i.e. a distance up to which 
such source should be detectable.  The rate with which a detector will
see mergers of a compact object binaries consisting of two
masses $M_1$ and $M_2$ with a lifetimes $t_{life}$ is
\begin{equation}
 {dR\over d{M_1 M_2}}=
\int_0^{z(M_1 M_2)} {dV \over dz } {f_{M_1 M_2}SFR(z+dz(t_{life}))\over 1+z} dz\, ,
\label{eqone}
\end{equation}
where $SFR(z)$ is the star formation rate function, 
$dz(t_{life})$ is the retardation due to the lifetime of the binary, 
and $f_{M_1 M_2}$ is the
probability of forming a binary with the masses $M_1$ and $M_2$,
while $z(M_1 M_2)$ is the sampling redshift out to which the detector is sensitive
to such mergers.
The population synthesis code provides an estimate of the function 
$f_{M_1 M_2}$, and gives us lifetimes of the binaries.
In the Euclidean case with constant star formation rate history 
this integration is elementary and yields \citep{2003ApJ...589L..37B}: 
\begin{equation}
{dR\over d{M_1 M_2}}\propto
{\cal M}^{5/2} \, .
\label{scal}
\end{equation}

In order to calculate the actual rate observed by a given detector one needs
a good estimate of the function $f_{M_1 M_2}$. A population synthesis
code provides an estimate of this function, i.e. the number of
such compact object binaries produced out of the initial population
considered by the code. One has to bear in mind that we do not simulate the 
entire stellar population, and a number of uncertain factors 
enters a calculation of the rate. First, we assume a binary fraction 
of $0.5$, yet this number could be different for real stellar population. 
Second the compact object binaries originate
in  stars at the high mass end of the initial mass function. 
We assume that the  the initial mass function is a power law, but 
its shape is not well known especially at the high end. The number 
of the compact object binaries depends strongly on the number of their 
progenitors. Only a small fraction of the total mass of the stars formed 
lies in the high mass stars. An additional uncertainty
may come from different low mass cutoffs of the initial mass 
function which affects the total number of stars considered.
The uncertainty in the rate from these factors can easily be  larger
than plus or minus a factor of ten. 
Additionally since the evolutionary times are 
important for the compact object
binaries one should know the   star formation history in our 
local group over that last few billion years.  
Apart from that, different models of stellar population lead 
different estimates of  $f_{M_1 M_2}$ \citep{2002ApJ...572..407B}.
The difference between the models when all other parameters
are kept constant is up to factor of a hundred between the most pessimistic
and most optimistic models.  
Thus given an observed rate will very hardly lead to strong
constraints on the stellar evolution model. Different models do predict
different rates when all the normalization parameters are kept constant.
However the uncertainty of  the  rates calculated 
within each model are large enough that imposing constraints
based solely on the rates  will be a difficult,
if not impossible task. Moreover, the  gravitational wave interferometers
exhibit non-stationary noise, which hampers   estimates of the 
times space volume sampled by a given detector.

\begin{figure}
\includegraphics[width=0.9\textwidth]{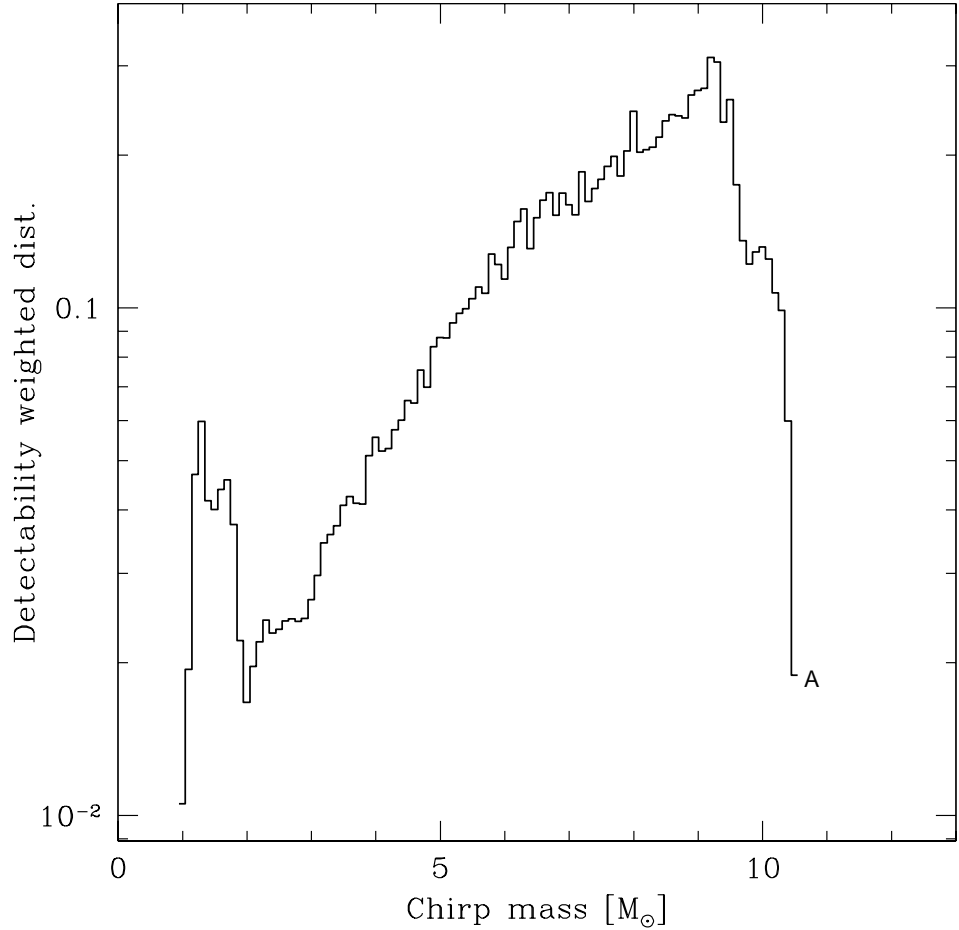}
\caption{The detectability weighted differential distribution of the 
chirp mass within the model A. Double neutron star mergers constitute
only less than a few percent of all the events.}
\label{chobs}
\end{figure}

Fortunately, future detections of gravitational waves from coalescing 
binaries will not produce the   rates alone. Each detection will be
accompanied by a measurement of the chirp mass of the coalescing binary,
and with good templates and post Newtonian analysis it should   be possible to
measure individual masses, and perhaps even the  spins. For the sake of this paper
we will only consider measurements of chirp masses. 
A galactic  population of compact object binaries is heavily dominated by 
small chirp mass objects, see Figure~\ref{chte}. However, the
detectability weighted distribution will be different. This is due to the 
fact the large chirp mass objects are visible to a larger distance.
Here we assume that the star formation rate history is given 
as in Figure~5 of  \citep{2002ApJ...571..394B}, i.e. it increases with redshift
by a factor of more than ten up to $z=2$ and then remains flat until
$z=30$.  We also assume a cosmological model with the Hubble constant
$H=65$km\,s$^{-1}$\,Mpc$^{-1}$, with $\Omega_m=0.3$ and $\Omega_\Lambda=0.7$,
and the model detector used for the purpose of this calculation is sensitive 
($S/N=8$) to 
a merger of a ${\cal M}=1.2\,M_\odot$ binary out to $125$\,Mpc.
We present the differential distribution of the
detectability  weighted distribution of chirp masses in Figure~\ref{chobs}.
We should note that a distribution of chirp masses as a
statistic is free from   nearly all of the problems that the rate has
suffered from. The distribution of chirp masses consists essentially 
of the ratios of one group of mergers to another. Therefore the 
absolute normalization  that has been such a problem in applying the 
rate calculation cancels out. The distribution of Figure~\ref{chobs} 
peaks at the high chirp mass objects. This is because the volume 
in which compact object coalescences   are detectable increases
rapidly with the chirp mass, see eq.~\ref{scal}.

\begin{figure}
\includegraphics[width=0.9\textwidth]{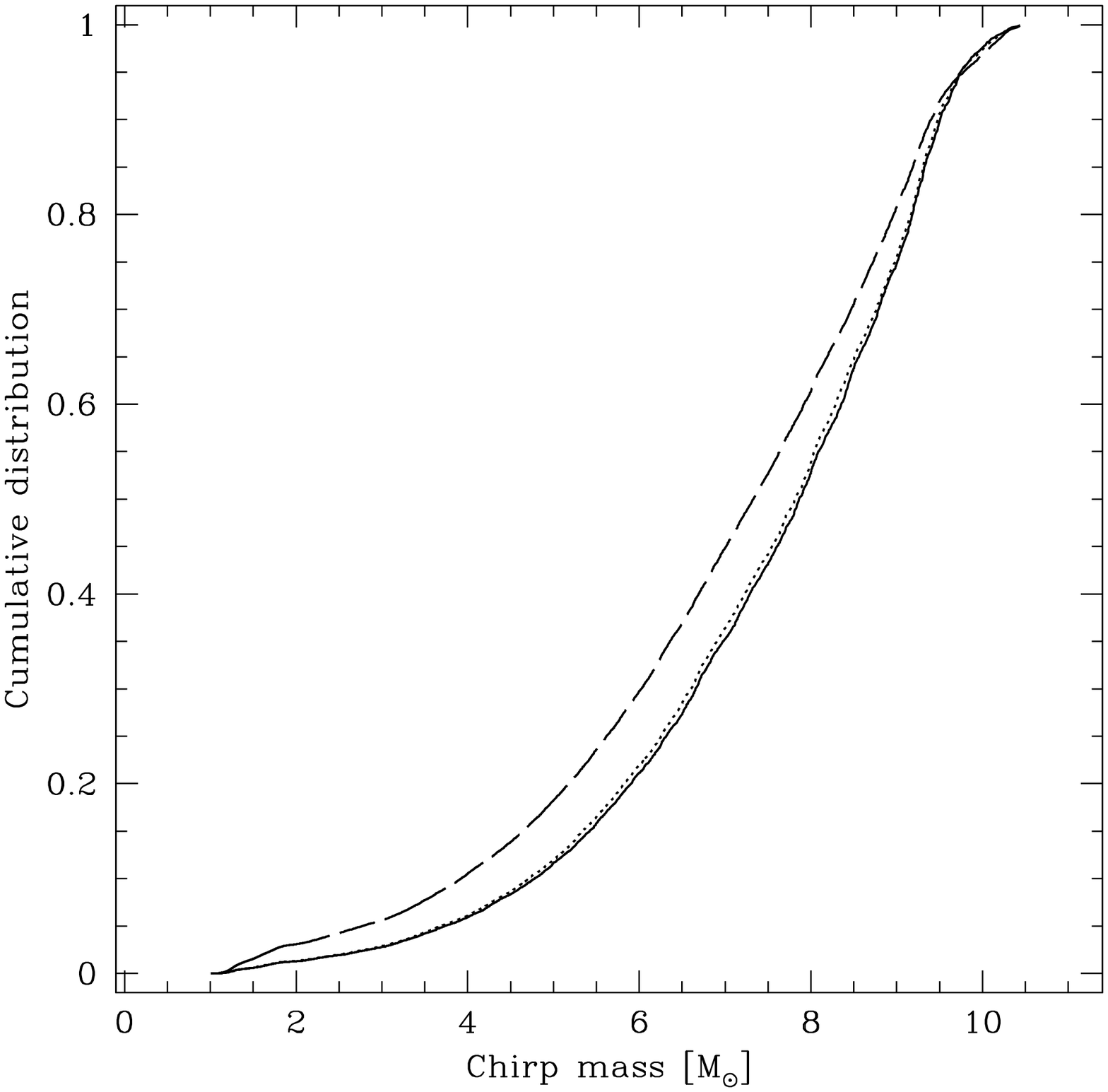}
\caption{The cumulative distributions of detectability weighted chirp masses.
the solid line corresponds to our fiducial model, while the dotted and dashed
lines
represent models where  detectors are more sensitive by a factor of ten and
hundred, respectively. }
\label{ddep}
\end{figure}

In the Euclidean space with  constant star formation rate the distribution of
observed chirp masses depends only on the assumed model of stellar evolution.
In Figure~\ref{ddep} we present how the detectability  weighted chirp mass
distribution changes when the sensitivity of the detector is increased, by a
factor of ten and hundred.  The differences are small, the distribution are
nearly the same  when the sensitivity  is increased by a factor of ten. The
increase of the sensitivity by a next factor of ten increases the number of low
chirp mass binaries in the observed sample. At this point the cosmological
effects play a role. The sampling distance to the  large chirp mass objects is
large, however there is not much volume at high redshifts. This increase of
sensitivity opens up  space for more detections of low chirp mass binaries, but
such detectors essentially see all double black hole mergers in the entire
Universe so increasing the sensitivity does not increase the number of high
chirp mass sources any more. A convenient method to compare two distributions
is to use the Kolmogorov-Smirnov test. This test uses a maximum distance
between the cumulative distributions $D$ to quantify the difference between 
them. The parameter $D$ between the solid and dotted line is very small - less
then $0.01$, while between the solid and dashed lines it is $\approx 0.07$. In
order to distinguish two distributions  at a significance level of about 
$10^{-4}$ one needs to sample  the distributions with  $N=4/D^2$ points.

\begin{figure}
\includegraphics[width=0.9\textwidth]{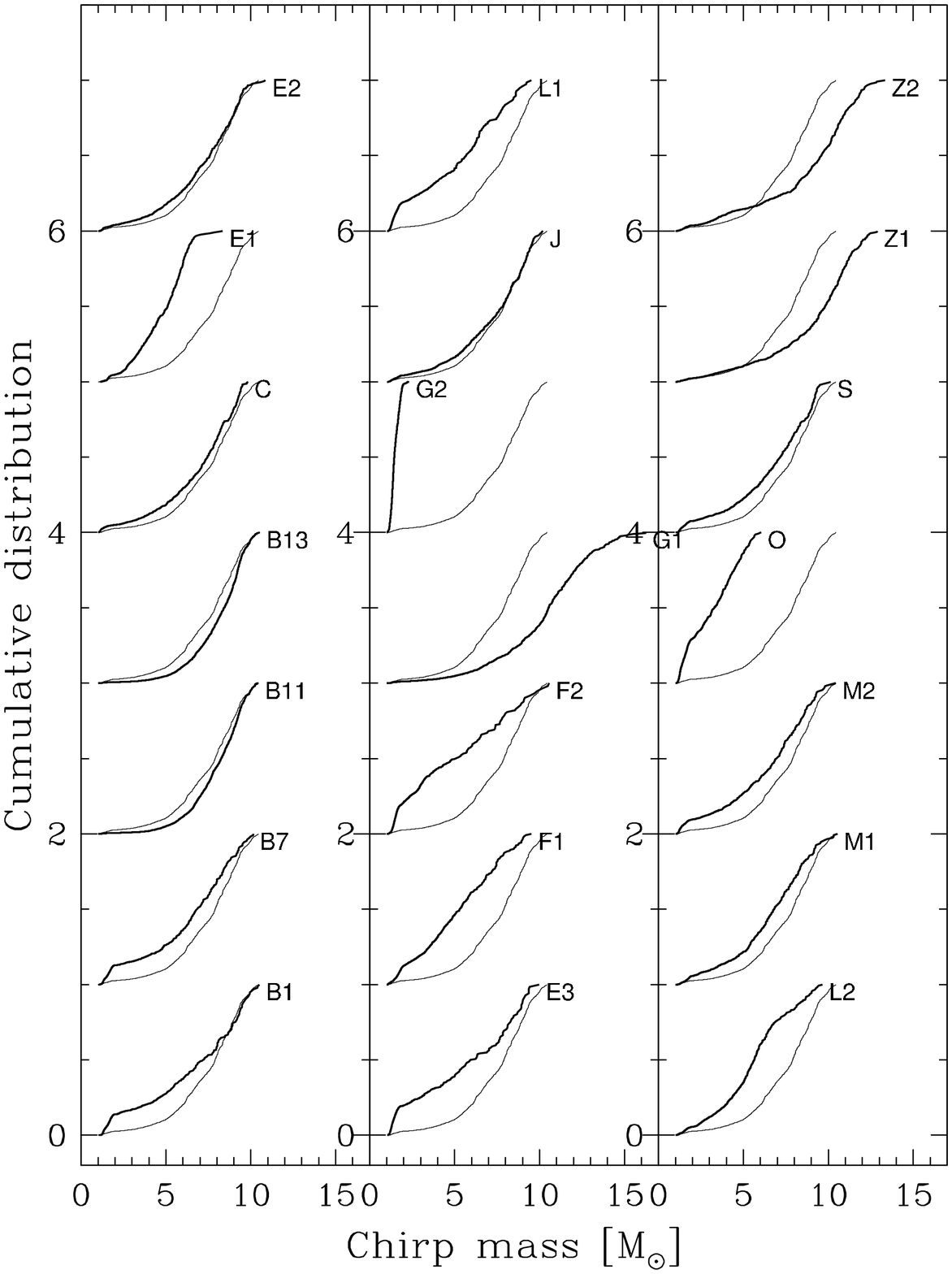}
\caption{Cumulative distributions 
of detectability weighted chirp mass within the models listed in Table~1. For
comparison we also show with thin lines 
 the distribution of model A .}
\label{four}
\end{figure}

In Figure~\ref{four} we present the cumulative distributions  of detectability
weighted chirp masses for the models listed in Table~1. For comparison we also
show in each panel the distribution corresponding to our model A. The
differences between the distributions are much larger than  these in
Figure~\ref{ddep} and for some models  the parameter $D$ is larger than $0.5$!
This shows that  a distribution of observed chirp masses carries a huge amount
of information about the underlying stellar evolution model.  It has been shown
\citep{2003ApJ...589L..37B} that most of the stellar evolutionary parameters
can be constrained with a sample of  one hundred merger observations.  Some
parameters affect the observed distribution of chirp masses even stronger:
several models can be ruled out with a sample of  less than 20 measured chirp
masses. Since the observed sample is dominated by the largest chirp mass
objects  the strongest constraints shall be imposed on these stellar
evolutionary parameters that affect production of massive black  holes. These
are in particular the details of the  core collapse of massive stars in
supernovae explosions, and also the averaged winds of most massive stars that
are the progenitors of highest mass black holes. It is important to  repeat
that a distribution of chirp masses is a statistic that is free of most of the
systematic uncertainties, which  constitute a problem, when one considers  just
the observed rate.

\section{Conclusions}

In this paper we presented the properties of the population of compact object
binaries relevant for the gravitational wave merger calculations.
The lifetimes of different types of binaries vary: the low mass binaries have
shorter
lifetimes than the heavy ones. The lifetimes of double black hole binaries 
are comparable to the Hubble time and therefore their progenitors
can originate in the epochs when star formation rate was much higher
than it is now. On the other hand the population of double neutron stars
is short lived and it originates in recent star bursts.

We discuss two possible observational statistics that can be used to constrain
stellar models: the observed rates, and the distribution of observed chirp
masses. We argue that the rates alone shall impose very weak constraints on the
stellar
models, because of huge uncertainties in the models, as well 
as possible difficulties in dealing with non stationary noise in the detectors.
The distribution of chirp masses is a statistics that is free from such
uncertainties. Most probably even stronger bounds would be imposed if
individual masses of coalescing objects are measured.

Finally we confirm the suggestion 
  \citep{1997NewA....2...43L} that the observed sample is dominated by 
 double  black hole binaries. In all the models that we consider 
 the observed sample is dominated by the highest chirp mass 
 objects.

\begin{theacknowledgments}
 This research was funded by the KBN grant 5P03D 011 20.
TB thanks the organizers of the meeting for support. 
\end{theacknowledgments}

\bibliographystyle{aipprocl} 

\newcommand{\apj}{Astroph. J.}
\newcommand{\apjl}{Astroph. J. Lett.}
\newcommand{\mnras}{M.N.R.A.S.}
\newcommand{\aap}{Astronomy and Astrophysics}
\newcommand{\prd}{Phys. Rev. D}

\end{document}